# El Uso de Repositorios y su Importancia para la Educación en Ingeniería

Texier, Jose[1,2,3]; De Giusti, Marisa[2]; Oviedo, Nestor[2]; Villarreal, Gonzalo L.[2,3]; Lira, Ariel[2]

jtexier@unet.edu.ve; {marisa.degiusti; nestor; gonzalo; ariel}
@sedici.unlp.edu.ar;

*1. Universidad Nacional Experimental del Tachira (UNET), Venezuela*
*2. Servicio de Difusión de la Creación Intelectual, Universidad Nacional de La Plata (SeDiCI), Argentina*
*3. CONICET, Argentina*

## Resumen

Los repositorios institucionales son depósitos de archivos digitales de diferentes tipologías para accederlos, difundirlos y preservarlos. Este artículo tiene como propósito explicar la importancia de los repositorios en el ámbito académico de la Ingeniería como una manera de democratizar el conocimiento por parte de los docentes, investigadores y alumnos para contribuir al desarrollo social y humano. Estos repositorios, enmarcados generalmente en la iniciativa Open Access, permiten asegurar el acceso libre y abierto (sin restricciones legales y económicas) a los diferentes sectores de la sociedad y de esa manera puedan hacer uso de los servicios que ofrecen. Finalmente, los repositorios están evolucionando en el ámbito académico y científico, y las diferentes disciplinas de la Ingeniería deben prepararse para brindar un conjunto de servicios a través de esos sistemas para la sociedad de hoy y del futuro.

**Palabras claves:** repositorios, ingeniería, open access, dataset, repositorios institucionales.

## Abstract

Institutional repositories are deposits of different types of digital files for access, disseminate and preserve them. This paper aims to explain the importance of repositories in the academic field of engineering as a way to democratize knowledge by teachers, researchers and students to contribute to social and human development. These repositories, usually framed in the Open Access Initiative, allow to ensure access free and open (unrestricted legal and economic) to different sectors of society and, thus, can make use of the services they offer. Finally, that repositories are evolving in the academic and scientific, and different disciplines of engineering should be prepared to provide a range of services through these systems to society of today and tomorrow.

# 1. Introducción

En los últimos años los Repositorios Institucionales (RIs) han cobrado importancia en la sociedad académica y científica, porque representan una fuente de información digital especializada, organizada y accesible para los lectores de diversas áreas. Los RIs son sistemas informáticos dedicados a gestionar los trabajos científicos y académicos de diversas instituciones de forma libre y gratuita, es decir, siguiendo las premisas del movimiento Open Access (OA).

El OA se entiende como el acceso inmediato a trabajos académicos, científicos, o de cualquier otro tipo sin requerimientos de registro, suscripción o pago. Por ello, este movimiento y los repositorios están ayudando a transformar el proceso de publicación de artículos científicos (1), permitiendo el acceso instantáneo o inmediato a las publicaciones arbitradas, gracias a las diferentes aplicaciones (Google Scholar, Microsoft Academic, Arxiv, Repositorios Institucionales de las Universidades) y servicios informáticos (alertas a partir de criterios previamente definidos, RSS, listas de correos, las diferentes redes sociales, etc.). Con el movimiento OA se han creado otros movimientos como el Open Data, Open Knowledge o Data Sharing, que incentivan el aumento de instalaciones y usos de los repositorios de documentos científicos y, en un porcentaje menor, de repositorios de documentos administrativos y de conjuntos de datos, conocidos también como dataset, raw data o datos crudos. En la Figura 1, se aprecia la curva de crecimiento de los repositorios registrados en el Directory of Open Access Repositories (OpenDOAR) (2), donde en diciembre del 2005 existían 28 repositorios registrados, y, en mayo del 2012 hay 2183 repositorios registrados. También, se puede extraer que la bandera "A" evidencia un crecimiento de 400 a 2183 repositorios, es decir, una incorporación de 1783 repositorios nuevos, representando un aumento de 445.75%. Para la bandera "B" el crecimiento es del 172.88% (se incorporaron 1383, de 800 a 2183), en la bandera "C" del 81.92% (se incorporaron 983, de 1200 a 2183), en la bandera "D" de 36.44% (se incorporaron 583, de 1600 a 2183) y en la bandera "E" de 9.15% (se incorporaron 183, de 2000 a 2183). Por tanto, se observa un crecimiento sostenido.

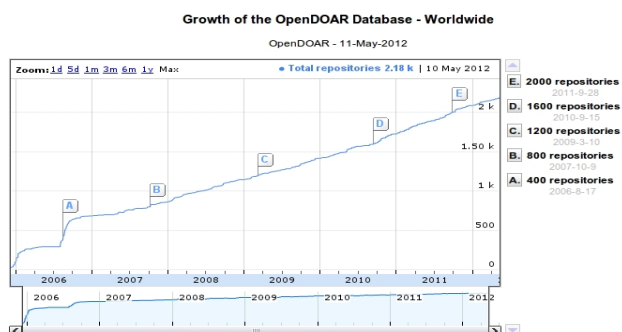

Fig. 1. Comportamiento de la Base de Datos del OpenDOAR

Nuestra investigación está centrada en revisar los repositorios de dataset en la Ingeniería y sus implicaciones educativas, fortaleciendo a los repositorios como una herramienta más en la formación del futuro ingeniero, además de lograr la preservación de esos documentos y datos en el tiempo, garantizando su acceso a futuras generaciones.

El Data Sharing consiste en compartir los datos de las investigaciones de los científicos, con el objetivo de unir esfuerzos y optimizar el uso de los recursos (3), es decir, que los data sets, puedan ser reutilizados con diferentes propósitos por el resto de los investigadores. Comenzó a estudiarse a partir de 1901, solicitando a los autores de publicaciones de estudios biométricos depositar los datos en algún lugar para su consulta (4). Luego, en 1971 inicia uno de los repositorios más antiguos, el Protein Data Bank. Después, surge la revista Journal of Biological Chemistry en 1983, que se convierte en una de las primeras revistas en exigir los datos de las investigaciones como condición para la publicación de artículos (5). Piwowar, en el 2007, muestra un incremento en el número de citas del 69% en revistas del área de la biomedicina (6). Estos hechos, establecen una percepción positiva en el uso de los repositorios en el área de la biomedicina pero no pareciera extenderse a las otras disciplinas, ya que los investigadores perciben un riesgo a publicar sus *data raw* y por lo que prefieren mantenerlas en sus computadoras con el riesgo que desaparezcan para siempre. Por ello, surge la duda de dónde depositar esos datos para poder preservarlos y compartirlos. Uno de los aportes de este trabajo, será contribuir con los investigadores en la Ingeniería para depositar sus trabajos.

En las estadísticas en OpenDOAR, muestran que solamente el 3.62% de repositorios (79 de 2183), permiten colocar dataset. Esta situación genera cierto desconcierto y desánimo en los investigadores para depositar los datos de sus investigaciones. Por tanto, es necesario incentivar la publicación a través del data sharing como práctica generalizada junto con políticas de creación y uso de los repositorios, ayudando al cambio cultural para una mejor formación del ingeniero. Un ejemplo es el editorial de Nature en 2009 (7) que expresa: "la comunidad científica, para llevar a cabo el *data sharing* necesita el equivalente digital de las bibliotecas actuales, es decir, alguien que preserve y haga accesible todos esos datos, apuntando directamente a las bibliotecas universitarias (como instituciones) y al data management (como rama del conocimiento) como los pilares sobre los que se debe apoyar el futuro del *data sharing*". Otros ejemplos similares se observan en otro editorial de Science (8), en diferentes artículos y políticas de publicación en revistas como British Medical, Plos ONE, el Archivo Nacional del Reino Unido, National Institute Health (NIH), Research Information Network (RIN), National Science Foundation, entre otras. Asimismo, diferentes instituciones universitarias, a través de sus bibliotecas, están avocadas a preservar los datos de los trabajos.

En este artículo se describe el concepto de los repositorios y la representación de los datos originales en las diferentes disciplinas de la Ingeniería. Luego, se exponen las implicaciones educativas de los repositorios, unas consideraciones finales y sugerencias para trabajos futuros.

## 2. LOS REPOSITORIOS
### 2.1 DEFINICIÓN
Los repositorios, también conocidos como repositorios digitales, están constituidos por un conjunto de archivos digitales en representación de productos científicos y académicos que pueden ser accedidos por los usuarios. Específicamente, los repositorios institucionales consisten en estructuras web interoperables de servicios informáticos, dedicadas a difundir la perpetuidad de los recursos científicos y académicos (físicos o digitales) de las universidades a partir de la enumeración de un conjunto de datos específicos (metadatos), para que esos recursos se puedan **recopilar, catalogar, acceder, gestionar, difundir y preservar** de forma libre y gratuita, por lo que están estrechamente ligados a los ideales y objetivos del Open Access. La representación de estos recursos se logra mediante el registro persistente del conjunto de datos asociados a ellos. Estos datos sirven como síntesis y reemplazo del objeto "real", lo cual permite distribuir el recurso sin requerir del objeto en sí, sino usando su representación. Las actividades de catalogación, acceso, gestión y difusión de los contenidos son las más consolidadas con el crecimiento de los repositorios, por el contrario, la recopilación de materiales y la preservación todavía se encuentran en sus primeros pasos.

A continuación se nombran los tres directorios más consultados según comScore.com (9), en los que se pueden observar algunas características de los repositorios registrados, a saber:

- OpenDOAR tiene registrados 2183 repositorios para el 13/05/12 (10).
- ROAR - Registry of Open Access Repositories tiene registrados 2875 repositorios para el 13/05/12 (11).
- University of Illinois OAI-PMH Data Provider Registry, con 2906 repositorios para el 13/05/12 (12).

Para este trabajo, se utiliza el OpenDOAR porque es el más consultado y sirve de fundamento para las estadísticas análizadas más adelante.

## 2.2 INICIATIVA DEL OPEN ACCESS

El Open Access (OA) tiene como fin asegurar el acceso libre y abierto a la producción científica. Una de las formas de lograr ese objetivo es por medio de la creación de repositorios institucionales donde se deposita esa producción científica para hacerla accesible sin restricciones y preservarla digitalmente como un bien común para la sociedad de hoy y del futuro. El movimiento de acceso abierto a la información se basa en dos estrategias fundamentales, una a través de las revistas de acceso abierto y la otra por medio de los repositorios institucionales. En 1966, se conoce el lanzamiento de Educational Resources Information Center (ERIC), biblioteca digital especializada en educación, y de Medline, una base de datos bibliográfica de biomedicina producida por la National Library Medicine (NLM) de los Estados Unidos. Una de las voces líderes es Peter Suber (13), que en el 2005 indica que existe una gran división en las publicaciones científicas, una referida a aquellas que están disponibles gratuitamente en la internet y otra en las cuales los lectores deben pagar para tener acceso a ellas. Además, enumera una serie de beneficios que ha generado, destacándose que los artículos en acceso abierto han sido citados 50-300% más que artículos que no están en OA en una misma revista, resaltando la importancia del autoarchivo como bandera del movimiento.

## 2.3 REPOSITORIOS DE DATOS

Los datos finales de investigación se entienden como el material factual registrado, aceptado por la comunidad científica y necesario para validar los resultados de la investigación, según el National Institutes of Health (3). Por tanto, no son datos finales de investigación: notas de laboratorio, sets de datos parciales, análisis preliminares, borradores de trabajos científicos, planes para investigaciones futuras, informes que han tenido un proceso de revisión por pares, comunicaciones con colegas, u objetos físicos como ejemplares de laboratorio. Torres-Salinas et al. en el 2012 (3), exponen diferentes taxonomías de los datos de investigación que sirven para determinar de una más clara qué se puede compartir y bajo qué contexto:

1. Según el formato
   - Textos
   - Números
   - Imágenes
   - Otros.
2. Proceso de obtención
   - Experimentales
     - Secuencias genéticas
     - Cromatografias
   - Simulaciones
     - Modelos climáticos
     - Modelos económicos
   - Observacionales
     - Encuestas
     - Experimentos irrepetibles.
3. Según objetivo recogida
   - Específicos
     - Sólo de interés para un proyecto de investigación
   - Alcance medio

- ○ De interés para una disciplina concreta.
- ● De interés general
  - ○ De interés para la ciencia en su conjunto e incluso de interés social.
4. Según fase de investigación
- ● Datos preliminares
  - ○ Datos recién extraídos sin ningún tipo de procesamiento. Denominados en inglés raw data
- ● Datos finales
  - ○ Datos que ya han sido procesados y combinados con otros (en inglés *final research data*.)

Esta clasificación sirve para categorizar inicialmente los repositorios, de esa manera se puede ayudar a la constitución de repositorios bajo un estándar. A continuación se nombran algunos proyectos que permiten confirmar el surgimiento de repositorios de datos:

| Proyecto o Trabajo | Características | Referencia |
|---|---|---|
| Lista Europea de Catálogos LOD2 | Se encuentran registrados 51 catálogos de Open Data de 27 países | http://lod2.okfn.org/eu-data-catalogues |
| Trabajo de Peter Kirlew | Repositorios de datos de Life science | http://www.istl.org/11-spring/refereed1.html |
| Universidad de Columbia | Repositorios de datos en: astronomía, ciencias biológicas, química, ciencias de la tierra y climáticas, datos marinos y ciencias sociales | http://scholcomm.columbia.edu/data-management/data-repositories/ |
| Narcis | Portal de repositorios de Holanda. Para el momento de la consulta: 24.207 datasets | http://www.narcis.nl/ |
| British University - Dryad | Datos de investigaciones revisadas por pares en biociencia. Tiene 1605 paquetes y 4018 archivos de datos. | http://www.datadryad.org/ |
| Cornell University - DataStar | Repositorio de datos experimentales | http://datastar.mannlib.cornell.edu/ |
| PDB/Protein Data Bank | Bioinformática | http://www.rcsb.org/pdb/home/home.do |
| GSA Data Repository | Geología | http://www.geosociety.org/pubs/drpint.htm |
| Cornell University Library | Física | http://arxiv.org/ |
| Chem,Seer | Química | http://chemxseer.ist.psu.edu/ |
| Organic Eprints | Agricultura | http://www.orgprints.org/ |
| GIS Data Resources | Geoespaciales | http://www.gis2gps.com/GIS/gisdata/gisdata.html |
| American Mathematical Society | Matemáticas | http://www.ams.org/global-preprints/ |
| SeaDataNet / NOAA | Datos marinos | http://www.seadatanet.org/ o http://www.nodc.noaa.gov/ |
| Repositories of Archaeology Data | Arqueología | http://openarchaeologydata.metajnl.com/repositories/ |
| DataCite | Instituto británico para el acceso de los datos de investigación | http://www.datacite.org/repolist |
| Australian Partnership for Sustainable Repositories | Centro para la gestión de datos académicos en formato digital | http://www.apsr.edu.au/ |
| Directorio mundial del Open Access | Lista de repositorios de datos | http://oad.simmons.edu/oadwiki/Data_repositories |

Tabla 1. Lista de Proyectos y Repositorios de Datos

Todos estos repositorios relacionados con datos fuentes para investigaciones, recopilados en la bibliografía relevada, indican que esta corriente se encuentra en un momento de plena expansión. Este hecho permite visualizar que tanto los investigadores en el área de las ciencias de la información como los administraciones de instituciones donde se prestan servicios relacionados, deben prepararse para auge de los repositorios de datos, que crecen a la par con la tecnología, situación que muchos autores llaman la era de la información (14).

## 2.4 Repositorios en la Ingeniería

En el 2012 Thompson clasifica a la Ingeniería en cuatro disciplinas principales o básicas (15): Ingeniería Química, Ingeniería Civil, Ingeniería Eléctrica e Ingeniería Mecánica. Luego el autor explica como las otras Ingenierías existentes son subdisciplinas e interdisciplinas (Ingeniería en Informática, Ingeniería de Materiales, Ingeniería Electrónica, Ingeniería Industrial, Ingeniería Agronómica, etc.). A partir de esta afirmación, se realiza una búsqueda de las disciplinas principales de la Ingeniería en el directorio de repositorios OpenDOAR, analizando los clasificados en las disciplinas básicas y los que tienen como contenidos a datasets.

La Figura 2, muestra las diferentes áreas representadas en el directorio. Se destaca que las cuatro disciplinas básicas, según Thompson, tienen 153 repositorios distribuidos con 63 en Ingeniería Química, 39 en Ingeniería Mecánica, 29 en Ingeniería Eléctrica y 22 en Ingeniería Civil. En la Figura 3, según los tipos de contenidos, los Journal Articles tienen 1467 repositorios (67.20%) y los Dataset tienen 79 repositorios (3.62%). Estos datos en concreto, permiten afirmar la poca presencia de esas disciplinas en los diferentes repositorios en el mundo, al igual que los data set.

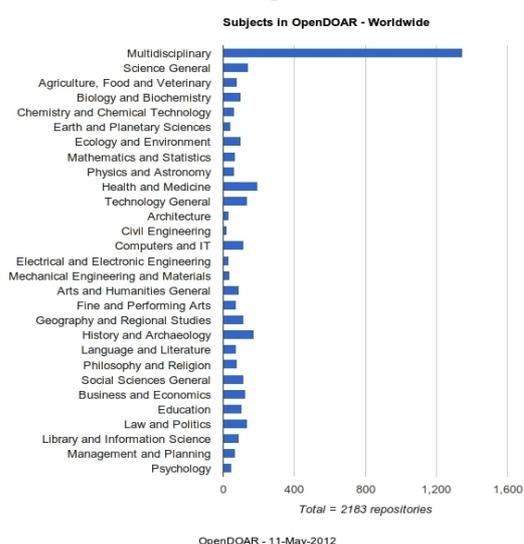

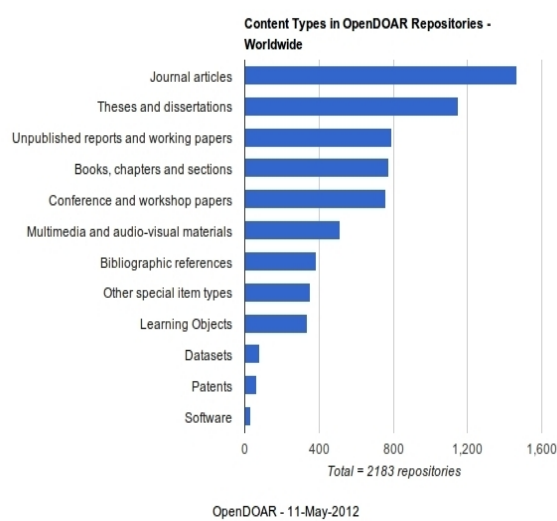

Fig. 2. Áreas en OpenDOAR    Fig. 3. Tipos de Contenido en OpenDOAR

La Tabla 2, expone la relación de los repositorios entre las cuatro disciplinas básicas y los dataset. Esta tabla se construyo a partir de una revisión sobre cuántos repositorios funcionaban o estaban activos. Se evidencia una debilidad en cuanto a número real de repositorios válidos, la poca existencia de políticas claras por cada una de estas disciplinas y la ausencia de estándares de metadatos, características propias de las entidades (papers, datasets, libros, etc), que permiten identificarlos para búsquedas, almacenamientos y recuperaciones. En cambio, las otras áreas que se mencionan a continuación, se observa que la comunidad de usuarios e instituciones afines han desarrollado estándares propios de metadatos, ayudando a consolidar los repositorios de datos:

- Darwin Core (Biology)
- DDI (Data Documentation Initiative for Social and Behavioral Sciences Data)
- DIF (Directory Interchange Format for Scientific Data)
- EML (Ecological Metadata Language)
- FGDC/CSDGM (Geographic Data)
- NBII (National Biological Information Infrastructure)
- MIAME (Minimum Information About a Microarray Experiment)
- MINSEQE (Minimum Information about a high-throughput SeQuencing Experiment).

| Clasificación | Total de Repositorios | Repositorios con Content Datasets | Porcentaje con Dataset | No Funciona | Porcentaje que no funciona |
|---|---|---|---|---|---|
| Ing. Civil | 22 | 2 | 9.09 % | 4 | 18.18 % |
| Ing. Química | 63 | 7 | 11.11 % | 21 | 33.33 % |
| Ing. Eléctrica | 29 | 0 | 0.00 % | 7 | 24.14 % |
| Ing. Mecánica | 39 | 4 | 10.26 % | 9 | 23.08 % |
| Totales | 79 | 15 | 18.99 % | 5 | 6.33 % |

Tabla. 2. Relación de las Disciplinas Básicas de Ingeniería y Datasets

Se estima que en los próximos años las diferentes disciplinas, subdisciplinas e interdisciplinas de la Ingeniería comenzarán a generar sus estándares de metadatos y de la misma forma también surgirán repositorios de datos. Por tanto, los ingenieros como las diferentes dependencias que están involucradas, deben prepararse para la consolidación de los repositorios en la Ingeniería y servicios que estimularán y cambiarán la educación en el área.

Los repositorios son sistemas que necesitan desarrollarse en alguna plataforma de software, por eso se observa en la Figura 4 la distribución de los repositorios registrados de acuerdo con el tipo de software usado. Se destaca que el líder es DSpace (desarrollado bajo licencia de software libre) con 872 repositorios con un 39.9%, a pesar de existir un 16.4% (358 repositorios) que están identificados por software desconocido, indicando que muchos de esos poseen desarrollos propios o simplemente no registraron su tipo de software. La tercera y cuarta posición son desarrollos en software libre: EPrints y Digital Commons. Entonces, se puede decir que las tres primeras plataformas de software para repositorios con licencias libres representan un 59.14%, es decir, 1291 repositorios.

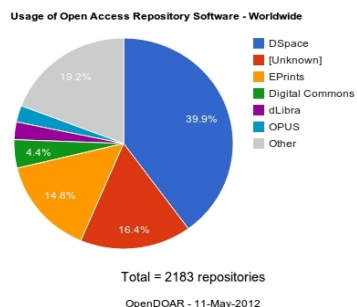

Fig. 4. Software de los Repositorios según OpenDOAR

## 2.5 PAUTAS PARA LA IMPLEMENTACIÓN DE REPOSITORIOS

Según el manual de LEADIRS II de Barton y Waters (16), se pueden tomar en cuenta los siguientes pasos para la implementación de repositorios:

- Aprendizaje sobre el proceso leyendo y examinando otros Repositorios Institucionales.
- Desarrollar una definición y un plan de servicio.
    - Realizar una evaluación de las necesidades de su Universidad.
    - Desarrollar un modelo de coste basado en este plan.
    - Crear una planificación y un horario.
    - Desarrollar políticas de actuación que gestionen la recopilación de contenidos, su distribución y mantenimiento.
- Formar el equipo.
- Tecnología (elegir e instalar el software).
- Marketing.
- Difundir el servicio.
- Puesta en funcionamiento del mismo.

Estas pautas son básicas y por supuesto pueden variar de acuerdo con cada institución, pero ayudan a marcar una línea lógica de implementación.

## 3. IMPLICACIONES EDUCATIVAS DE LOS REPOSITORIOS

Dave Turek, el encargado del desarrollo de supercomputadoras de IBM, declaró en mayo del 2012, que desde el comienzo de la historia humana hasta el año 2003, se ha generado según cálculos de IBM, cinco exabytes, es decir, cinco mil millones de gigabytes de información. El año pasado, se generaban esa misma cantidad de datos cada dos días y para el próximo año, pronóstica Turek, se generará lo mismo cada 10 minutos (17). Esta afirmación definitivamente es una evidencia de que se vive en la era de la información, la misma que se consolidó gracias al desarrollo de internet y de las tecnologías de la información (14). Este cambio de época se puede observar en lo cultural, social, económico y tecnológico, por ello, hoy día los niños y jóvenes crecen con un computador o dispositivo electrónico, conocidos como nativos digitales (18).

La situación descrita por Turek plantea a la sociedad, empresas e instituciones académicas un gran reto relacionado con la gestión de grandes conjuntos de datos así como con la generación de información. La educación, en todos los niveles, debe comenzar a pensarse con base en la época en la que se vive, permitiendo entre otras cosas el acceso libre y gratuito a todos los datos a través de repositorios, los cuales garantizan la recopilación, difusión y preservación de la información para la sociedad de hoy y del futuro. Gran parte de esos datos provienen de la investigación que debe cumplir con los estándares de calidad, por ejemplo, la revisión por pares. Por lo general, los trabajos se apoyan en trabajos anteriores y dependen principalmente de las posibilidades que tengan los científicos de consultar y compartir las publicaciones científicas y los datos de la investigación. Por tanto, la difusión de la información debe ser rápida para que pueda contribuir a la innovación y evitar la repetición de investigaciones con aporte a la ciencia.
En el 2011, Castro et al. (19) muestran en su investigación cómo los repositorios están transformando la educación primaria y secundaria en Portugal, cambiando poco a poco todas las estructuras y formas de pensar de los actores en esos niveles, favoreciendo la calidad académica. De igual manera, Xia y Opperman (20) en el 2009 destacan la importancia de los repositorios en la formación de futuros magister y grados intermedios, y reportan que el 49.50% de los recursos depositados pertenecen a trabajos estudiantiles, conviertiéndolos en actores principales de los nuevos trabajos disponibles en los repositorios. Ambos trabajos concluyen: que los contenidos digitales provenientes de diversas fuentes están aumentando, la existencia de intercambios de esos contenidos, la publicación de esos contenidos en repositorios, la reutilización de la información se realiza todo el tiempo y los software de esos repositorios están en su mayoría, con licencias de software libre. A continuación se nombran algunas potencialidades y limitaciones (19) para fundamentar la relación entre repositorios y la educación:

*- Potencialidades:*
- Facilitar la modificación de las prácticas pedagógicas.
- Fomentar las prácticas de enseñanza más interactiva y constructiva.
- Inducir y facilitar la producción y utilización de herramientas, contenidos, recursos e información en formato digital.
- Facilitar enfoques de colaboración en la enseñanza.
- Minimizar la brecha digital, permitiendo el acceso remoto y contenidos de bajo coste, módulos y cursos.
- Fomentar la inclusión en la enseñanza y el aprendizaje de los ciudadanos con necesidades especiales.
- Desarrollar y fortalecer una cultura de aprendizaje permanente.
- Mantener la información en el tiempo y garantizar su acceso a próximas generaciones.

*- Limitaciones:*
- Técnica: falta de disponibilidad de internet en algunos sectores.
- Económico: la falta de recursos para invertir en hardware y software, limitando el desarrollo de herramientas informáticas y mantenimiento de proyectos a largo plazo.
- Social: la ausencia de habilidades para utilizar las invenciones técnicas.

- Cultural: resistencias en la distribución o el uso de los recursos producidos por otros profesores o instituciones.
- Políticas de estado y marcos legales.

Finalmente, el compartir (data sharing) los datos primarios o data raw de las investigaciones, permitirá aumentar la eficiencia de la investigación y la calidad de la educación que recibirán los estudiantes, ya que esos dataset se pueden utilizar para explorar las nuevas hipótesis o las relacionadas, además de ser indispensables para el desarrollo y la validación de los métodos de estudio, técnicas de análisis e implementaciones de software. Por supuesto, que si esos datos están libres y gratuitos para los estudiantes, se generarán mayor innovación en ellos -entre otros beneficios-, ayudando a identificar los errores en etapas tempranas de los trabajos y evitando la recolección de datos duplicados. Por tanto, parte del éxito de la educación estará en que los datos se puedan recopilar, catalogar, acceder, gestionar, difundir y preservar, es decir, se necesitan y se necesitarán repositorios para la Ingeniería.

## 4. CONSIDERACIONES FINALES

- Se requieren más estudios sobre el comportamiento de los repositorios de datos en la Ingeniería, ya que el acceso a la información científica, su difusión y su preservación constituyen importantes retos en la era digital.
- En el trabajo se evidencia la debilidad existente en la formalización de repositorios en la Ingeniería, pero simplemente se observa como un llamado de atención para continuar ayudando a construir estándares y herramientas que permitan la fácil adaptación a las diversas disciplinas de la Ingeniería.
- El movimiento de Data Curation esta tomando una gran importancia en el mundo de los repositorios, ya que permitirá enriquecer la comprensión de difundir y preservar datos. Existen proyectos como los coordinados por la University of Illinois (21) y Purdue University (22), o cursos en postgrados (23 y 24) para fomentan que los materiales deben ser calidad y garantiza el acceso a largo plazo.
- Incrementar la visibilidad de la producción académica y científica de las Ingenierías, a partir de la consolidación e incremento de los repositorios en las universidades, tomando en cuenta, la poca cantidad de repositorios en esas instituciones, además de la falta de sitios web de las diferentes disciplinas de las ingenierías, permitiendo conocer con más detalle esas disciplinas.
- Partiendo del principio que no existen políticas claras entre los investigadores acerca de cómo conservar sus datos, ya que muchas veces los datos están dispersos en diferentes medios de almacenamiento (CDs, DVDs, discos duros externos, PCs, correos, en la nube), esto provocará pérdidas irrecuperables por diversas causas, por ello deben establecerse políticas que ayuden a la comunidad a conservar sus datos en el tiempo.
- Las revistas deben continuar obligando que en las publicaciones se exijan que los datos utilizados se almacenen en el repositorio, por tanto, los repositorios deben contar con la plataforma acorde para permitir catalogar todas las diversidades de datos existentes. Por ejemplo: Nature, Science, Plos one, etc.
- Los repositorios y los Learning Management Systems (LMS) tienen objetivos completamente distintos, los repositorios están pensados para el acceso, difusión y preservación de documentos y datos, en cambio, las plataformas e-learnings integra un conjunto de herramientas para la enseñanza-aprendizaje en línea, de forma no presencial o mixta, generando una interacción alumno-profesor.
- Una recomendación para los formatos de los archivos de los datos que se quieren conservar es que sus formatos no sean propietarios, que no estén comprimidos y que no estén encriptados.

## 5. TRABAJOS FUTUROS

- Establecer políticas para la constitución y catalogación en los repositorios de datos.

- Generar una propuesta de ranking propio que tome en cuenta características de la disciplina en la que se encuentre, por ejemplo las estadísticas de colecciones de data set de economía llamado RePEc (25).
- Analizar los modelos de los repositorios referentes o los más consultados para proponer un modelo general de repositorios, permitiendo generar otras plataformas o desarrollar componentes sobre los ya existentes, usando metodologías como MDA o líneas de productos de software.

## Referencias